# Fundamental Magnetic Properties and Structural Implications
# for Nanocrystalline Fe-Ti-N Thin Films

Jaydip Das, Sangita S. Kalarickal, Kyoung-Suk Kim, and Carl E. Patton

*Department of Physics, Colorado State University, Fort Collins, Colorado 80523*

The magnetization ($M$) as a function of temperature ($T$) from 2 to 300 K and in-plane field ($H$) up to 1 kOe, room temperature easy and hard direction in-plane field hysteresis loops for fields between ±100 Oe, and 10 GHz ferromagnetic resonance (FMR) profiles have been measured for a series of soft-magnetic nano-crystalline 50 nm thick Fe-Ti-N films made by magnetron sputtering in an in-plane field. The nominal titanium concentration was 3 at. % and the nitrogen concentrations ($x_N$) ranged from zero to 12.7 at. %. The saturation magnetization ($M_s$) vs. $T$ data and the extracted exchange parameters as a function of $x_N$ are consistent with a lattice expansion due to the addition of interstitial nitrogen in the body-centered-cubic (bcc) lattice and a structural transition to body-centered-tetragonal (bct) in the $6-8$ at. % nitrogen range. The hysteresis loop and FMR data show a consistent picture of the changes in both the uniaxial and cubic anisotropy as a function of $x_N$. Films with $x_N \geq 3.9$ at. % show an overall uniaxial anisotropy, with an anisotropy field parameter $H_u$ that increases with $x_N$. The corresponding dispersion averaged uniaxial anisotropy energy density parameter $\langle K_u \rangle = H_u M_s / 2$ is a linear function of $x_N$, with a rate of increase of $950 \pm 150$ erg/cm$^3$ per at. % nitrogen. The estimated uniaxial anisotropy energy per nitrogen atom is 30 J/mol, a value consistent with other systems. For $x_N$ below 6 at. %, the scaling of coercive force $H_c$ data with the sixth power of the grain size $D$ indicate a grain size averaged effective cubic anisotropy energy density parameter $\langle K_1 \rangle$ that is about an order of magnitude smaller that the nominal $K_1$ values for iron, and give a quantitative $\langle K_1 \rangle$ vs. $D$ response that matches predictions for exchange coupled random grains with cubic anisotropy.

## I. INTRODUCTION

The nanocrystalline Fe-Q-N thin film system, with Q = Al, Ti, Ta, or Zr, for example, has been a subject of great interest over the last decade or so.[1-10] This interest has been driven, in part by the seminal work on binary Fe-N films by Kim and Takahashi,[11] who were the first to show that the addition of nitrogen appears to give a magnetization higher than pure iron, even at small nitrogen levels. While the nitrogen is generally taken to reside on interstitial sites and result in an expansion of the iron lattice,[12] the actual effect of the nitrogen atoms on the magnetization is still a matter of controversy.[13-15] The shift in interest to the three element Fe-Q-N film system has been driven by the realization that small amounts of the third element can provide an enhanced thermal stability.[16, 17] In addition to the challenges to understand the fundamental magnetic properties of Fe-Q-N films, the system shows a number of attractive properties for microwave device and magnetic information storage applications. These include large magnetization, field induced uniaxial anisotropy, high permeability, low coercive force, and low microwave loss.

The recent work by Alexander and co-workers on the Fe-Ti-N film systems, as cited above, has shown evidence for a low microwave loss with strong microstructure correlations.[8, 9] Work from this group has also provided concise data on structure, room temperature magnetization, and anisotropy.[5-7, 10] Apart from these studies, however, there has been very little definitive work to elucidate the magnetic interactions, identify structural transitions and changes in the magnetic properties associated with these transitions, or to separate the different contributions to the magnetic anisotropy.

The purpose of this work was to perform a comprehensive study of the fundamental magnetic properties of the Fe-Ti-N film system. This has been done for a range of nitrogen concentrations from $0-13$ at. %, the range over which most of the interesting and useful property changes occur. Three types of measurements were made, magnetization vs. temperature from 2 to 300 K at a fixed field of 1 kOe, hysteresis loops for fields from $-100$ to $+100$ Oe at 300 K, and room temperature field swept ferromagnetic resonance (FMR) at 9.5 GHz. All data were for in-plane fields only.

The different types of data all indicate some sort of a structural transition in the $6-8$ at. % nitrogen range. This applies to the effective exchange as extracted from the magnetization vs. temperature data, the coercive force data, and the cubic and uniaxial anisotropy parameters obtained from the hysteresis loop and the FMR data. There is a clear transition from cubic anisotropy dominance to uniaxial anisotropy dominance in the same $6-8$ at. % nitrogen range.

The paper is organized as follows: Section II provides a brief overview of the structure and magnetic properties of the Fe-N and Fe-Q-N film systems. Section III gives a description of the film materials and the magnetization and microwave measurement procedures. Section IV presents magnetization vs. temperature data and an analysis of the data in terms of effective exchange parameters. Section V presents the room temperature microwave and hysteresis loop data, summarizes results on the coercive force, and shows a qualitative analysis of the data in terms of cubic and uniaxial anisotropy. Section VI elucidates the dependence of derived cubic and uniaxial anisotropy energy density parameters on nitrogen content, based on the data presented in Sec. V, and examines the ferromagnetic exchange length implications for these nanocrystalline films.

## II. OVERVIEW OF THE Fe-N and Fe-Q-N FILM SYSTEMS

The introduction gave a brief overview of Fe-N and Fe-Q-N thin films, with an emphasis on magnetic properties. This section focuses on structure. For the film thicknesses of interest here, typically in the 50 nm range, no special thickness effects are expected and only bulk structure considerations are needed. The available literature indicates that as one adds nitrogen, there are progressive changes in structure.[18] One starts with body-centered-cubic (bcc) $\alpha$ – Fe and a saturation induction $4\pi M_s$ of about 21 kG. For $Fe_{1-x_N} N_{x_N}$, with a nitrogen content $x_N$ up to 0.4 at. % or so, the nitrogen atoms can be dissolved in the bcc lattice with essentially no change in the structure.[18, 19] It is important to emphasize that the nitrogen enters the bcc lattice interstitially and the $x_N$ – parameter here simply denotes the atomic fraction of the nitrogen in the overall material.

As one moves above $x_N \approx 0.4$ at. %, the interstitial nitrogen becomes sufficient to produce a tetragonally distorted $\alpha'$– Fe-N phase. At some level between $x_N \approx 0.4$ at. % and $x_N \approx 11$ at. %, the literature indicates an overall structure change to a body-centered-tetragonal (bct) $\alpha''$– Fe-N phase.[18] Reported values of the saturation induction of the $\alpha''$ phase vary over a wide range, typically from 20 to 30 kG or so.[13-15, 17] Neither the details on this bcc-bct transition point nor the problem with the wide variance in $4\pi M_s$ have been resolved to date. Above $x_N \approx 11$ at. %, there appears to be a transition to a face-centered-cubic (fcc) $\gamma'$– Fe-N phase with the nitrogen atoms at interstitial octahedral sites. At about $x_N \approx 25$ at. %, one further structural change gives rise to a hexagonal-closed-packed $\varepsilon$ – Fe-N phase. Finally, for $x_N \approx 50$ at. %, the orthorhombic $\varsigma$ – Fe-N phase forms.[12]

Even though the sequences of changes summarized above produce interesting phase transitions and potentially useful magnetic properties in some regimes, the metastability of these binary Fe-N phases[18] limits the practical use of these films. It has been found that the addition of small amounts of a third element Q to form ternary Fe-Q-N phases, typically Hf, Ta, Ti, or Zr, can serve to increase thermal stability.[4, 16, 17] For pure iron, additions of such a Q element at levels up to about 5 at. % have been found to dissolve substitutionally into the bcc lattice with no substantial effect on the structure or magnetic properties.[3] Similar levels of substitution for these elements in the Fe-N system also appear to have little effect except for the enhanced thermal stability. This is one reason for the focus on Fe-Ti-N films in this work. The distorted bcc to bct structural transition noted above for the Fe-N film system also appears to be operative for Fe-Ti-N films. Data in Ref. 6, for example, indicate that films with about 3 at. % Ti, have a structural transition at about $x_N \approx 7$. at. %.

By way of example, Fig. 1 shows schematic crystal structure diagrams for (a) bcc iron and (b) tetragonally distorted bcc iron with an interstitial nitrogen atom in place, as indicated. The shaded circle in each diagram serves to indicate the possible presence of iron site titanium atoms in the overall Fe-Ti-N structure. The labels indicate nominal atom spacings. The non-parentheses and parentheses values in (a) are for pure bcc iron and Fe-Ti with about 3 at. % titanium, respectively. The values in (b) are estimates for Fe-N with $x_N \approx 5$ at. %, under the assumption that there is no phase transition to bct at this level.

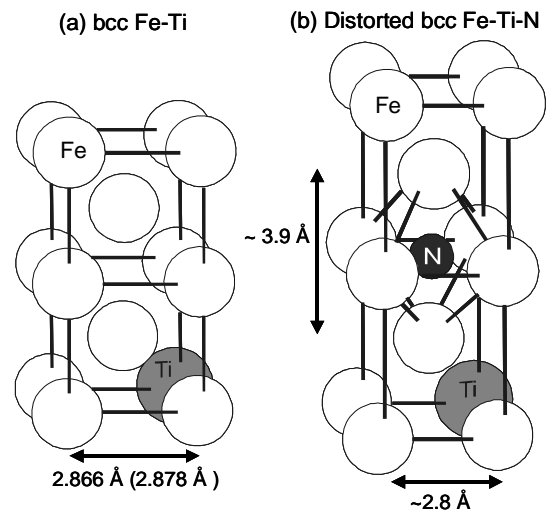

FIG. 1. Schematic illustrations of (a) the bcc iron structure and (b) the tetragonally distorted bcc Fe-N structure for about 5 at. % nitrogen. The shaded atoms indicate possible titanium additions. The separation distances in (a) are for pure iron (non-parentheses) and iron with about 3 at % titanium (parentheses). The separation distances in (b) are estimated values for Fe-N with about 5 at. % nitrogen and no structural transition to bct.

Overall, Fig. 1 shows two things. First, the addition of nitrogen to the bcc Fe or Fe-Ti system causes a tetragonal distortion to the lattice. Second, the titanium, if present, resides on iron sites. The choice to place a single interstitial nitrogen atom on the lattice diagram in Fig. 1 (b) is simply for purposes of illustration. As noted above, one can have a tetragonally distorted bcc structure at low $x_N$ levels or a bonafide body-centered-tetragonal structure for high $x_N$ levels. Insofar as no unit cell is indicated, the (b) diagram could actually apply to either of these situations. The main point of Fig. 1 (b) is that the addition of nitrogen breaks the cubic symmetry. One should notice that the interstitial nitrogen in the (b) diagram serves to push away the two nearest-neighbor iron atoms. It is this symmetry breaking effect that can lead to a uniaxial magnetic anisotropy in field deposited alloys or films.

The additional point to be drawn from the figure concerns the bcc site titanium. It was noted above that the Fe-N system with interstitial nitrogen is generally unstable and that the presence of bcc site titanium atoms serves to stabilize the Fe-N system. This is believed to occur because the increase in the lattice parameter caused by the larger titanium atom increases the space available for the interstitial nitrogen. The data show that for small levels of Ti, this makes for a more stable alloy.

The role of the nitrogen in modifying the magnetic anisotropy in Fe-Ti-N films is one of main aspects of this paper. The body-centered-cubic Fe-Ti alloy has cubic anisotropy. As nitrogen is added to the lattice, there is a decrease in the cubic anisotropy, up to the structural transition to bct at $x_N \approx 7$ at. %.[7] The nitrogen induced distortion in the lattice might be one possible reason for this decrease. Above the $x_N \approx 7$ at. % structural transition point, the cubic anisotropy is essentially zero.

There is no literature data available on the effect of nitrogen on the field induced uniaxial anisotropy in Fe-Ti-N films. For Fe-Ta-N and Fe-Zr-N films, deposition in an in-plane field produces a uniaxial anisotropy that increases with $x_N$.[21, 22] One can presume that this is a pair ordering effect. One can also speculate that Fe-Ti-N films should show a similar response. It seems clear that the symmetry breaking due to the nitrogen, first in producing a tetragonally distorted bcc lattice at low $x_N$ levels and then in producing a bonafide bct structure at higher levels, should produce such an anisotropy when a field is used to promote a preferred direction in the overall polycrystalline film.

One of the key results from the present work is a clear map of the change in the anisotropy, both cubic and uniaxial,

as a function of the nitrogen level in the Fe-Ti-N system. Based on the above, one would expect to see a drop in the cubic anisotropy to zero as one goes from Fe-Ti to Fe-Ti-N with $x_N \approx 7$ at. %. At the same time, one should see an increase in the uniaxial anisotropy component, and this anisotropy will become dominant as the alloy changes to bct for $x_N > 7$ at. % or so. One can also speculate that these anisotropy changes will be accompanied by signature changes in the magnetization. Data from the literature do show that the magnetization drops as $x_N$ is increased.[5] There has been no work that demonstrates any connection between the structural change induced effects on the anisotropy with related effects on the magnetization. The data given below will demonstrate such connections.

In addition to structural changes, the presence of nitrogen affects the grain-growth process in Fe-Ti-N films. Transmission electron microscopy data on films similar to those used in this work indicate a decrease in grain size from about 28 nm at $x_N = 0$ to 4 nm at $x_N \approx 12.7$ at. %.[8, 9] This effect is relevant to the present work because the hysteresis loop properties depend on microstructure as well as anisotropy. References [8] and [9] also show a connection between grain size and the FMR linewidth. However, it is not clear that the linewidth effect is related to grain size or to a structural change. The hysteresis loop data below will make a case for correlations with both structure and grain size.

## III. FILM PREPARATION AND MEASUREMENT TECHNIQUES

The samples were kindly provided by Professor C. Alexander, Jr. of the University of Alabama MINT Center. The film preparation and the materials analysis work was done by Dr. Y. Ding and Professor Alexander. The nano-crystalline Fe-Ti-N films of nominal 50 nm thickness were prepared on glass substrates in a nitrogen-argon atmosphere by DC magnetron sputtering. A 300 Oe static magnetic field was applied in the plane of the substrate to define a specific axis for the possible development of the field induced uniaxial anisotropy. For purposes of discussion this direction will be termed the "easy" direction, even for films with no uniaxial anisotropy. The films were then removed from the deposition system and annealed at 100 °C in an easy axis oriented 300 Oe field. The atomic concentrations of the different elements were measured by x-ray photoelectron spectroscopy (XPS). All films had about 3 at. % of titanium. The nitrogen content $x_N$ varied from zero to 12.7 at. %. Grain sizes were determined by transmission electron microscopy (TEM). The nitrogen content and average grain sizes of various films are listed in Table I. Further details are given in Refs. [6] and [8].

TABLE. I.  Summary of sample parameters as a function of nitrogen content.  The nitrogen content and grain size for different films were obtained from XPS and TEM measurements at the MINT Center, University of Alabama.  Other parameters were obtained from SQUID and FMR measurements as part of this work.  These parameters are discussed in detail in Secs. IV and V.

| Nitrogen content $x_N$ (at. %) | Grain size $D$ (nm) | Saturation induction $4\pi M_s$(300 K) (kG) | Spin-spin exchange energy $J_{ex}$ ($\times 10^{-15}$ erg) | Coercive force $H_c$(300 K) (Oe) | Hard axis hysteresis loop determined uniaxial anisotropy field parameter $H_u$(300 K) (Oe) | FMR Uniaxial anisotropy field parameter $H_u$(300 K) (Oe) |
|---|---|---|---|---|---|---|
| 0 | 28 | 19.4 | 6.7 | 13.5 | - | - |
| 1.9 | 20 | 19.0 | 4.4 | 7.5 | - | - |
| 3.9 | 15 | 18.8 | 3.7 | 6.7 | 7 | 4.0 |
| 5.4 | 10 | 18.2 | 3.0 | 6 | 9 | 6.5 |
| 7.0 | 8.5 | 16.8 | 2.5 | 5.3 | 11 | 8.7 |
| 8.4 | 7.5 | 15.3 | 2.4 | 6.5 | 16 | 12.5 |
| 10.9 | 5 | 14.2 | 2.1 | 8.5 | 19 | 12.5 |
| 12.7 | 4 | 13.9 | 2.1 | 11.0 | 23 | 19.0 |

The magnetization measurements were made as a function of temperature and magnetic field with a superconducting quantum interference device (SQUID) magnetometer.  The fields were always applied in-plane.  Prior to the SQUID measurements, FMR measurements as a function of the in-plane field direction were used to ascertain the easy and hard axes for the films with uniaxial anisotropy.  The details of the FMR measurements are given below.  Three types of magnetization data were then obtained.  First, full magnetization vs. field hysteresis loop data were obtained for easy direction in-plane fields from −100 to +100 Oe at 300 K.  Fields of 100 Oe were found to be adequate to achieve magnetic saturation in all cases.  Second, saturation magnetization vs. temperature was then measured from 300 K down to 2 K at a fixed field of 1 kOe.  All of the magnetization values in Table I and the magnetization data below are based on a nominal film thickness of 50 nm for all the samples.

Finally, in order to determine the threshold nitrogen level needed for uniaxial anisotropy and determine the uniaxial anisotropy field values for samples above this threshold, additional hysteresis loop data were obtained at 300 K for both easy and hard direction fields.  Samples with $x_N \geq 3.9$ at. % showed uniaxial anisotropy.  Those with $x_N < 3.9$ at. % showed no uniaxial anisotropy.  The coercive force values from the easy direction hysteresis loop measurements were also used to estimate values of the cubic and uniaxial anisotropy energy density parameters based on established coercive force models.  Anisotropy nomenclature will be introduced as needed in Secs. V and VI.

Room temperature ferromagnetic resonance measurements at 9.5 GHz were used to check the uniaxial anisotropy field values for the high nitrogen content samples.  The FMR profiles were measured by a shorted waveguide reflection technique[23, 24] with field modulation and lock-in detection.  The samples were mounted on the waveguide short such that the external static and microwave fields were in the plane of the films and mutually perpendicular.  The raw data consisted of plots of the field derivative of the FMR absorption peak response as a function of applied field.  The resonance field position was determined from the zero crossing point on the derivative profile.  Data were obtained as a function of the in-plane static field orientation in 10 degree steps over a half circle.  The FMR field position vs. angle was then used to obtain the uniaxial anisotropy field parameter $H_u$.  For films with $x_N < 3.9$ at. %, the derivative FMR profiles were too broad and the shifts in the FMR field position with angle were too small to obtain usable values of $H_u$.

Table I shows a summary of basic film parameters as a function of the nitrogen level.  The nitrogen content and grain size shown in the first two columns were obtained from XPS and TEM measurements by Professor C. Alexander, Jr. of the University of Alabama MINT Center.[25]  The other columns show results from the SQUID and FMR measurements.  The saturation induction and spin-spin exchange energy results will be considered in Sec IV.  The coercive force and uniaxial anisotropy field results will be considered in Sec. V.

## IV.  EFFECT OF NITROGEN ON SATURATION INDUCTION

As noted in the introduction, the initial interest in Fe-N thin films was driven, in large part, by an apparent increase in the magnetization for small additions of nitrogen.  Interestingly, there has been little work on the change in the magnetization with the nitrogen level.  The use of titanium to achieve thermally stable films makes the current series of films ideal for a full study of the magnetization as a function of the nitrogen content as well as temperature.

There are two parts to this section. The first focuses on the actual data on the saturation induction $4\pi M_s$ vs. temperature $T$ for the full series of samples. The second presents an analysis of the $M_s(T)$ response, sample by sample, to give the $x_N$ dependence of the average nearest-neighbor spin-spin exchange interaction for the Fe-Ti-N films. The analysis shows a strong correlation between the expected structural changes in the films with increasing $x_N$ and the exchange. One finds, specifically, that the Heisenberg exchange parameter $J_{ex}$ shows a very strong decrease with increasing $x_N$ up to about $6-8$ at. % nitrogen and then shows only a very small change, if any, for larger $x_N$ values. This response is consistent with the expected expansion of the lattice with the addition of nitrogen and a structure change at $6-8$ at. %.

### A. Nitrogen loading and temperature dependence of saturation induction

Figure 2 shows experimental results on the saturation induction vs. temperature and nitrogen level. The data for each sample were obtained as a decreasing function of temperature and for a constant in-plane field $H$ of 1 kOe. Graph (a) shows the full ensemble of data in a $4\pi M_s(T)$ format for all of the samples, with the $x_N$ values as indicated. Graph (b) shows selected data in $4\pi M_s$ vs. $x_N$ format at two specific temperature points, 50 K and 300 K. The solid curves in (a) represent fits to a Bloch $T^{3/2}$ law. The curves in (b) are a guide to the eye.

Figure 2 (a) shows four things. First, at a given $x_N$ level, $4\pi M_s$ is a decreasing function of temperature. Second, the $4\pi M_s(T)$ response can always be fitted nicely to a Bloch $T^{3/2}$ type response. Third, the curvature of the $4\pi M_s(T)$ response generally increases with $x_N$. This is an indication that the exchange coupling is a decreasing function of $x_N$. Finally, at any given temperature, $4\pi M_s$ generally decreases with increasing nitrogen content. There is never an increase in $4\pi M_s$ with $x_N$.

Figure 2 (b) elaborates on the last point made above. These data show that the $4\pi M_s$ vs. $x_N$ response at fixed temperature shows an interesting structure. In particular, the 50 and 300 K data both show a gradual decrease in $4\pi M_s$ up to about $x_N = 6$ at. %, followed by a rather rapid change in the $x_N = 6-8$ at. % range, and then a gradual leveling off for higher $x_N$ values. It is also useful to take note of the numerical values in graph (b). First, the $4\pi M_s$ value at 300 K and $x_N = 0$ is about 19.5 kG. This value is very close to the $4\pi M_s$ value of 21 kG for pure $\alpha - \mathrm{Fe}$ noted above. The small difference is attributed to the nonmagnetic titanium in these films. Second, one can see that there is very little scatter in the point-to-point data for either temperature. Keep in mind that all of the data are for different films with assumed nominal thicknesses of 50 nm.

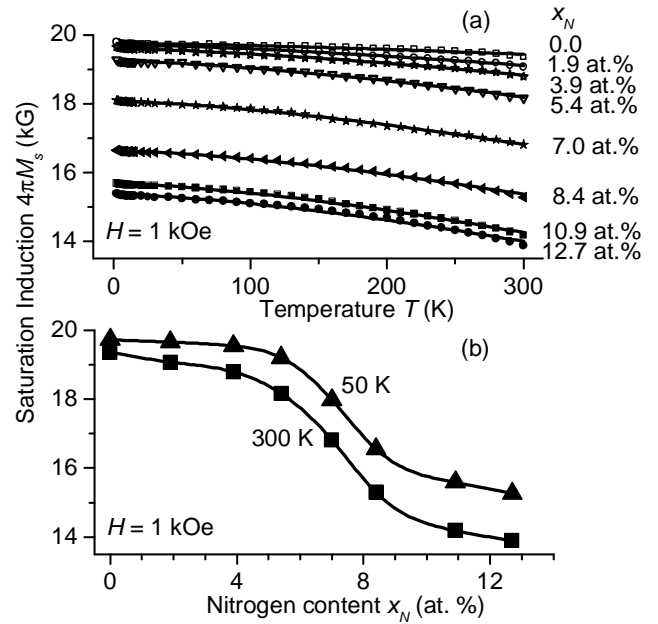

FIG. 2. (a) Saturation induction $4\pi M_s$ as a function of temperature $T$ for films with different nitrogen concentrations ($x_N$), as indicated. The data were obtained for an in-plane field $H$ of 1 kOe and taken as a decreasing function of temperature. The solid lines show Bloch $T^{3/2}$ law fits. (b) Variation in $4\pi M_s$ with $x_N$ at 50 K and 300 K. The curves provide a guide to the eye only.

The lack of scatter indicates that all of the films have the same thickness to within a few percent or so. This means that overall, the $4\pi M_s$ data in Fig. 1 show the true relative response for the different $x_N$ values.

Turn now to the Bloch $T^{3/2}$ law fits in Fig. 2 (a). Within the framework of the Heisenberg exchange spin-wave model for a three-dimensional ferromagnetic system at temperatures well below the Curie temperature, the temperature dependence of the saturation induction can be written as[26]

$$4\pi M_s(T) = 4\pi M_{s0}[1 - BT^{3/2}].  \qquad (1)$$

Here, $4\pi M_{s0}$ is the extrapolated saturation induction at $T = 0$ and $B$ is the Bloch coefficient. The $B$ parameter allows one to test the experimental response against the spin-wave model and quantify the curvature of the $4\pi M_s(T)$ response. As noted above, the solid line fits to the Bloch law in Fig. 2 (a) are quite good. The $R^2$ (correlation coefficient squared) of the fit for the $x_N = 0$ film is equal to 0.96 and for all other films, this value is above 0.99. These fits, in turn, yield empirical values of $4\pi M_{s0}$ and $B$ as a function of $x_N$. The fitted $4\pi M_{s0}$ as a function of $x_N$ shows the same basic response as the Fig. 2 (b) data at 50 K, but with values that are slightly above the points shown.

The fitted $B$-values range from $(3.1 \pm 0.4) \times 10^{-6}$ K$^{-3/2}$ at $x_N = 0$ to $(17 \pm 1) \times 10^{-6}$ K$^{-3/2}$ at $x_N = 12.7$ at. %. The overall $B(x_N)$ response is essentially linear up to $x_N = 10.9$ at. % and then appears to level off. The value for $x_N = 0$ closely matches literature value for bulk $\alpha$-Fe, $B_{\alpha-Fe} = 3.3 \times 10^{-6}$ K$^{-3/2}$.[27] The increase in the $B$-parameter with $x_N$ is an indication of a decrease in the average exchange interaction between the spins. A quantitative analysis of the $B(x_N)$ results in terms of exchange will be considered below.

Finally, consider the $4\pi M_s$ vs. $x_N$ responses in Fig. 2 (b). The decrease in $4\pi M_s$ with increasing $x_N$ at a given temperature can be attributed to the magnetic dilution caused by more and more nitrogen on interstitial sites. One can surmise, moreover, that the rapid change in $4\pi M_s$ in the $x_N = 6-8$ at. % range is indicative of some sort of a structural transition. Such a transition is, in fact, reflected in the exchange response as a function of the nitrogen content. The following subsection considers the actual exchange energy parameter $J_{ex}$ as a function of $x_N$.

## B. Variation in the spin-spin exchange energy with nitrogen content

The standard starting point for a spin-wave analysis for $M_s(T)$ is a mean-field nearest-neighbor Heisenberg hamiltonian of the form,

$$\mathcal{H} = -2J_{ex} \sum_i \mathbf{S}_i \cdot \mathbf{S}_{i+1}, \qquad (2)$$

where the $\mathbf{S}_i$ denote individual lattice site spins. From the standard Heisenberg analysis,[26] one obtains an $M_s(T)$ as given in Eq. (1) with a $J_{ex} - B$ connection given by

$$J_{ex} = \left( \frac{k_B}{2S} \right) \left[ \frac{SQB}{0.0587} \right]^{-\frac{2}{3}}. \qquad (3)$$

In the above, $k_B$ is the Boltzmann constant, $S$ is the dimensionless spin value for the site, and $Q$ is the number of atoms per cubic unit cell in the lattice. The numerical factor derives from the boson statistics for spin waves. For bcc iron, one has $S = 5/2$ and $Q = 2$. This isotropic mean-field model is taken to be applicable for polycrystalline films of the sort used here. The important point for this discussion is that $J_{ex}$ varies as $B^{-2/3}$. This means that the linear increase in $B$ with $x_N$ noted above will translate into a non-linear decrease in $J_{ex}$ with $x_N$.

One final caveat is needed. Equation (3) is explicitly for a bcc structure. In accord with the $4\pi M_s(x_N)$ data in Fig. 2 (b), the $J_{ex}$ results indicate the possibility of a structure transition that will, in turn, invalidate the specific form of Eq. (3). This point will be revisited at the end of the section.

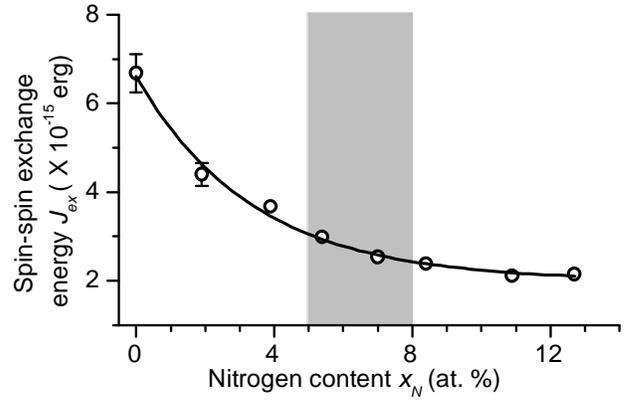

FIG. 3. Spin-spin exchange energy $J_{ex}$ as a function of nitrogen content $x_N$. The open circles show the application of the Heisenberg model to the measurement results. The solid line shows a best fit based on an exponential decay function of the form $U + V e^{-W x_N}$ with $U = 2 \times 10^{-15}$ erg, $V = 4.6 \times 10^{-15}$ erg, and $W = 0.3$. The shaded region for $5 < x_N < 8$ at. % indicates a possible structural transition region. The errors for the points without error bars are smaller than the size of the symbol.

Figure 3 shows the change in $J_{ex}$ with $x_N$ from the data fits to Eq. (1) and the conversion in Eq. (3). An exponential fit, as specified in the caption, is shown by the solid curve. The gray shading in the $5 < x_N < 8$ at. % range is shown to indicate a possible structural transition. Apart from the good exponential decay response, the overall drop in $J_{ex}$ from $x_N = 0$ to 12.7 at. % is about 70 %.

In line with the discussion above, the overall decrease in $J_{ex}$ with increasing $x_N$ can be attributed to the increase in the spin-spin distance as the added interstitial site nitrogen causes the lattice to expand. The strong decrease in $J_{ex}$ for $x_N < 6$ at. % or so, indicates an increase in the average spin-spin distance in this range. For higher concentrations, the apparent leveling off in $J_{ex}$ indicates that some sort of transition has occurred. As noted in Sec. II, there is a structural transition somewhere in the $0.4 < x_N < 11$ at. % range for Fe-N, and it is reasonable to expect a similar effect for Fe-Ti-N. X-ray data by Ding *et al.*,[6] on films similar to those studied here do indicate a structural change.

What do these data mean? After the discussion of Sec. II, it is clear that the random distribution of the nitrogen atoms at the octahedral interstices first distorts the bcc structure for relatively low $x_N$ levels and then results in a bcc to bct structural transition at large $x_N$. Recall that the $4\pi M_s(x_N)$ data in Fig. 2 (b) also indicate a possible structure transition in the $6 < x_N < 8$ at. % range. Based on these considerations and the results in Fig. 3, the overall $J_{ex}(x_N)$ response can be explained as follows. (1) At small values of $x_N$, likely below 6 at. % or so, there is a dilation in the overall bcc lattice due to an increase in the Fe site-to-site distance induced by the presence of the interstitial nitrogen, and this dilation causes $J_{ex}$ to decrease rapidly. (2) As $x_N$ moves above 6 at. % or so, the structural transition suggested in Ref. [6] and evident from Fig. 2 (b) causes an apparent leveling off in the $J_{ex}(x_N)$ response. As noted above, however, a structural transition invalidates the rigorous use of Eq. (3) to extract a $J_{ex}$ from the fitted Bloch coefficient $B$. This means that the numerical values of $J_{ex}$ in Fig. 3 for large $x_N$ levels may not be strictly applicable.

As an aside, it is useful to recast the exchange parameter $J_{ex}$ in terms of the often-used exchange stiffness constant parameter $A$. For a bcc lattice, the $A - J_{ex}$ connection is given by $A = 2J_{ex}S^2/a$, where $a$ is the lattice constant.[28] The $A$-format is used extensively and one can readily compare present results with values in the archival literature. Typical $A$-values for many materials, from metals to insulators, are in the $10^{-6}$ erg/cm range. For $\alpha$-Fe, one has $A = 2.1 \times 10^{-6}$ erg/cm,[29] for example. For Fe-Ti films, Ding *et al.*[10] report an $a$ value of 2.878 Å. Based on the $J_{ex}$ results in Fig. 3, this gives $A = (2.9 \pm 0.2) \times 10^{-6}$ erg/cm at $x_N = 0$.

# V. COMPETITION BETWEEN CUBIC AND UNIAXIAL ANISOTROPY

This section focuses on the effect of the nitrogen level on the cubic and the field induced uniaxial anisotropy. Both ferromagnetic resonance and hysteresis loop results are considered. Nitrogen is known to affect the magnetic anisotropy in Fe-N and Fe-Q-N films. The bcc $x_N = 0$ film has a predominant cubic magnetocrystalline anisotropy and no measurable field induced uniaxial anisotropy. As $x_N$ is increased from zero and nitrogen goes to the interstitial sites, there is a local breaking of the cubic symmetry. This symmetry breaking might cause a reduction in the usual cubic anisotropy. At the same time, the interstitial nitrogen, in combination with film deposition in a field, gives rise to a uniaxial anisotropy. Further, it is well known that the grain size governs the anisotropy in the nanocrystalline systems. As discussed earlier, the nitrogen also serves to slow the grain growth in the Fe-Ti-N system. This gives the grain size decrease with increasing nitrogen content shown in Table I.

There are two parts to this section. Part A presents room temperature FMR results that show a clear signature of a deposition field induced uniaxial anisotropy. The uniaxial field parameter increases with $x_N$ for nitrogen levels above about 4 at. %. Part B presents room temperature hysteresis loop results. The easy direction loop based coercive force data and the hard direction loop anisotropy data for $x_N \geq 3.9$ at. % reveal systematic changes in both the cubic and uniaxial anisotropy with nitrogen content as well as a competition between these anisotropies. Both results are consistent with the considerations outlined above. Section

VI will provide a semi-quantitative analysis of these results in terms of energy density considerations.

## A. Microwave results

Measurements of the fixed frequency FMR field $H_{FMR}$ as a function of the angle $\theta$ between the in-plane applied field and the easy axis can be used to determine the uniaxial anisotropy field parameter $H_u$. For iron based films and x-band frequencies, the operational FMR formula that connects $H_{FMR}$, $\theta$, and $H_u$ can be written as[30]

$$\omega_{FMR} = |\gamma| \sqrt{4\pi M_s (H_{FMR} + H_u \cos 2\theta)}, \quad (4)$$

where $\omega_{FMR}$ is the resonance frequency and $\gamma$ is the electron gyromagnetic ratio for the material. For most transition metal alloys, one can make the pure spin approximation and use a nominal spin only $|\gamma|/2\pi$ value of 2.8 GHz/kOe. It is important to emphasize that Eq. (4) is valid only if two conditions are satisfied, (1) a saturated film and (2) $4\pi M_s \gg H_{FMR}$. Subject to these conditions, fits from Eq. (4) to data on $H_{FMR}$ as a function of $\theta$ can be used to determine $H_u$. The data show that for the films with $x_N \geq 3.9$ at. %, there is generally a clear in-plane uniaxial anisotropy for these field deposited films. As noted in Sec. III, the FMR linewidths for films with $x_N < 3.9$ at. % were too broad to obtain a discernable $H_{FMR}(\theta)$ response or usable values of $H_u$.

Figure 4 shows room temperature microwave results for samples with $x_N \geq 3.9$ at. %. Graph (a) shows representative FMR resonance position vs. angle data for one particular film with $x_N = 8.4$ at. %. The open circles show the data and the solid curve shows a best fit from Eq. (4). The curve shown is for $4\pi M_s = 15.3$ kG, as obtained from the room temperature saturation induction measurements discussed in Sec. IV and an $H_u$ value of 12.5 Oe. The $R^2$ value for the fit was 0.99. Graph (b) shows the full ensemble of microwave results on $H_u$ as a function of $x_N$, based on data and fits similar to those in (a). The solid line shows a linear fit to the data shown. Films with $x_N < 3.9$ at. % showed no measurable uniaxial anisotropy.

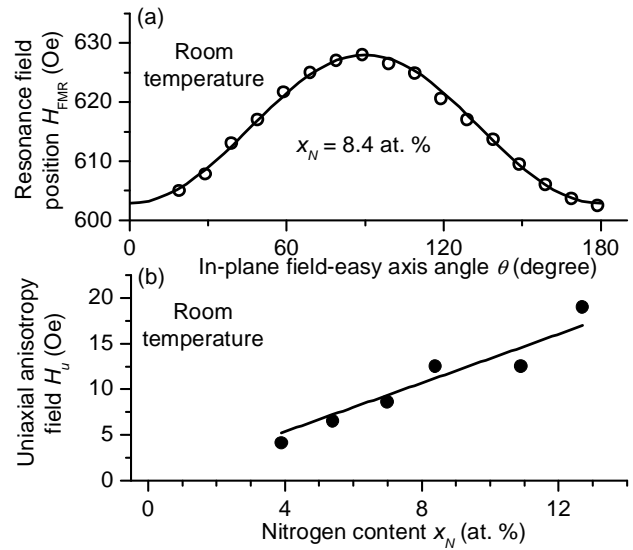

FIG. 4. (a) Room temperature FMR resonance position $H_{FMR}$ as a function of the angle $\theta$ between the in-plane field and easy direction for the $x_N = 8.4$ at. % film. The solid curve shows the fit to Eq. (4) with $4\pi M_s = 15.3$ kG and $H_u = 12.5$ Oe. (b) Uniaxial anisotropy field parameter $H_u$ as a function of nitrogen content $x_N$. The solid line shows a linear fit to the data.

The results in Fig. 4 show that (1) the FMR field vs. in-plane easy axis-field angle data fit nicely to a uniaxial anisotropy model, and (2) the uniaxial anisotropy field parameter $H_u$ appears to increase with $x_N$ for $x_N \geq 3.9$ at. %. The response is approximately linear. The slope of the fitted line in Fig. 4 (b) is 1.3 Oe/at. %. This increase is consistent with a picture of localized tetragonal distortions to the lattice induced by nitrogen additions and a directional ordering of these local regions for field deposited films.

It is important to emphasize that these FMR determinations of $H_u$ vs. $x_N$ are done for saturated samples. In contrast with the hysteresis results in the next subsection, models based on magnetization processes do not enter into the results.

## B. Hysteresis loop results

Hysteresis loop measurements as a function of the field angle also provide information on anisotropy. Ideally, easy direction in-plane fields will give square loops with a coercive force $H_c$. One can use easy direction $H_c$ data to estimate either the uniaxial anisotropy energy density parameter $K_u$ or the cubic anisotropy energy density parameter $K_1$, depending on which one is dominant. If the cubic anisotropy plays the dominant role in the magnetization processes, hard direction hysteresis loops will be almost similar to the easy direction loops. If, however, uniaxial anisotropy plays the dominant role, the ideal hard direction loop response will take the form $M = (M_s / H_u) H$ for $H < H_u$ and $M = M_s$ for $H > H_u$. Here, $M$ is the magnetization and $H$ is the static external field. For non-ideal uniaxial systems with, for example, interactions between misaligned grains, these ideal hard direction uniaxial loops will be somewhat widened.

The hysteresis loops look quite different for the films with $x_N < 3.9$ at. % and those with greater amounts of nitrogen. Figure 5 shows representative 300 K easy and hard direction in-plane field hysteresis loops for samples below and above this cut. The data are shown in a normalized $M(H)/M_s$ format. Graph (a) is for $x_N = 1.9$ at. % and (b) is for $x_N = 7$ at. %. The solid and dashed lines in each graph show the easy and hard direction loop data, respectively. The loops in (a) are quite square and nearly identical, and with a coercive force of about 7.5 Oe. The fact that the sample shows no evidence of uniaxial anisotropy is consistent with the FMR results.

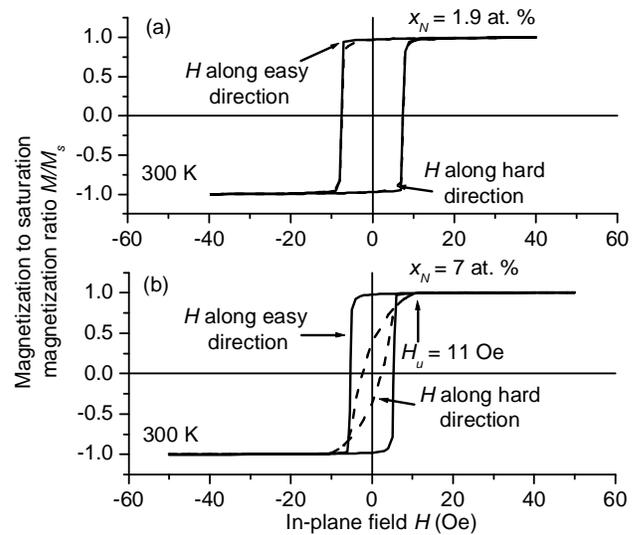

FIG. 5. Example room temperature hysteresis loops of the magnetization $M$, normalized to the saturation magnetization $M_s$, as a function of the in-plane field $H$ for two Fe-Ti-N films. Graphs (a) and (b) are for nitrogen contents ($x_N$) of 1.9 at. % and 7 at. %, as indicated. The solid and dashed lines show the data for easy and hard direction fields, as indicated. The hard direction saturation field indicated in the first quadrant of (b) is taken as the uniaxial anisotropy field parameter $H_u$ for the sample.

The loops in (b), on the other hand, have a clear uniaxial anisotropy signature. The saturation point for hard direction loop in (b) is taken as the $H_u$ parameter for the sample. The easy direction loop is nearly square and the coercive force is about 5 Oe. The saturation point at 11 Oe noted for the hard direction loop is consistent with the 9 Oe value for $H_u$ obtained by FMR. One can also note that the hard direction loop in (b) has an open character. This is an indication of a fairly high degree of anisotropy dispersion for these polycrystalline films.

Figure 6 shows summary data from the hysteresis loop measurements for all samples. The solid square data in (a) show results on the coercive force $H_c$ values from the easy direction loops as a function of the nitrogen content $x_N$. The inset shows the same $H_c$ data, but plotted as a function of the average grain size $D$ from Table I. The solid curve in the inset shows a fit to a model function $H_c(D) = X + Y D^6$. Following Herzer and co-workers,[31, 32] a $D^6$ response would be expected for a system of randomly oriented single-domain interacting cubic grains. Keep in mind that the large grain size end of the inset graph corresponds to low nitrogen and the low grain size region corresponds to high nitrogen. These $H_c$ vs. $D$ data and the fit shown will be important for the discussion below.

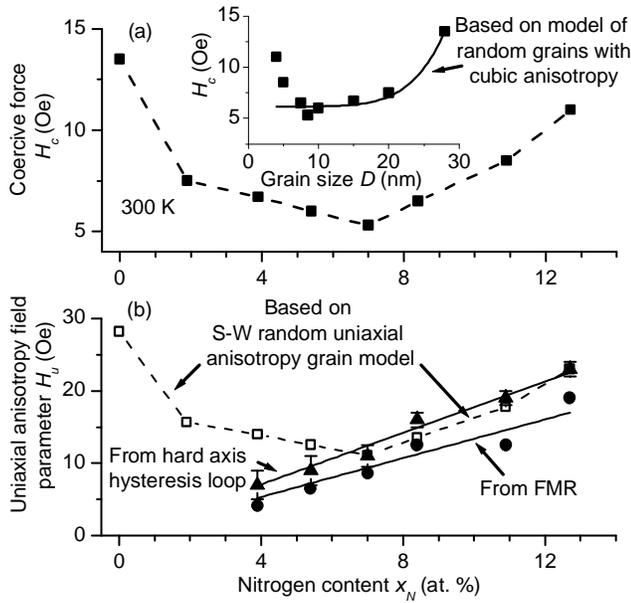

FIG. 6. (a) Coercive force $H_c$ as a function of nitrogen content $x_N$ at 300 K. The solid squares show the data. The dashed lines are a guide to the eye. The inset shows $H_c$ as a function of grain size $D$. The solid line shows a fit of the form $X + YD^6$, with $X = 6$ Oe and $Y = 1.53 \times 10^{34}$ Oe/cm$^6$. (b) Uniaxial anisotropy field parameter $H_u$ as a function of $x_N$. The solid triangles show the $H_u$ values obtained from the hard axis hysteresis loop data. The solid circles show the FMR based $H_u$ values from Fig. 4 (b). The solid lines show linear fits to these two data sets. The open squares show inferred $H_u$ values based on the $H_c$ data in (a) and a Stoner-Wohlfarth model for randomly oriented non-interacting uniaxial grains. The dashed lines connect the points.

In Fig. 6 (b), the solid triangles show the $H_u$ vs. $x_N$ results from the hard direction loops for the samples with $x_N \geq 3.9$ at. %. The solid circles show the corresponding $H_u$ vs. $x_N$ results from the FMR measurements. The solid straight lines represent linear fits to the respective sets of data. The open square points and the dashed lines show the $H_u$ vs. $x_N$ prediction based on a Stoner-Wohlfarth (S-W) random uniaxial grain model.[33] This model suggests a coercive force to anisotropy field conversion recipe given by $H_c = 0.479\,H_u$.

In Fig. 6 (a), one sees that the coercive force decreases with increasing nitrogen content for low $x_N$ values, goes through a minimum for $x_N \approx 7$ at. % film, and then increases for higher concentrations. This type of response has been confirmed by separate measurements by the Alabama group[25] and is also found for Fe-Zr-N films.[22] The same data in an $H_c$ vs. $D$ format, and the fit to the random cubic grain model, serve to illustrate the possible cubic anisotropy origin of the $H_c$ vs. $x_N$ response at low nitrogen levels and the need for another mechanism at high $x_N$. The fit shows that the data for $7.5$ nm $\leq D \leq 28$ nm and $8.4$ at. % $\geq x_N \geq 0$ matches such a model, and that there are significant deviations from the predictions of a random cubic grain coercive force model for smaller grain sizes and higher nitrogen levels.

Figure 6 (b) provides complimentary information based on uniaxial anisotropy considerations. The data show good consistency between the hard axis hysteresis loop saturation field $H_u$ points for $x_N \geq 3.9$ at. % and the FMR results. The up shift for the hysteresis loop data can be attributed to the choice of the hard direction loop saturation field as a measure of $H_u$. The slope of the fit to the loop results is about $1.8 \pm 0.1$ Oe/at. %. This consistency is noteworthy, insofar as the loop data involve magnetization processes that take a high dispersion sample from a partially magnetized state to a saturated state, while the FMR data are for saturated samples at high field.

The $H_u$ vs. $x_N$ response shown by the open square points and connecting lines in graph (b), estimated on the basis of the S-W randomly oriented uniaxial anisotropy model, follow the similar trend as the $H_c$ data in (a). The more interesting aspect is that these points appear to match the actual $H_u$ data reasonably well for $x_N \geq 7$ at. %. For lower $x_N$ values, this model gives a rapidly increasing $H_u$ prediction that is completely different from the data trend.

These data provide several previously unrealized connections between nitrogen content and anisotropy in Fe-Ti-N films. Keep in mind that an increasing $x_N$ level has two effects. It produces an emerging uniaxial anisotropy for $x_N$ values above 4 at. % or so. At the same time, the induced tetragonal distortion is also expected to ameliorate the sizeable cubic anisotropy that is present for the pure Fe-Ti system. The main graph in Fig. 6 (a) shows both effects nicely. The drop in $H_c$ with increasing $x_N$ in the left side of the graph is due to a *decrease* in the cubic anisotropy. The increase in $H_c$ with $x_N$ on the right side is due to the *increase* in $H_u$. The inset results reinforce this scenario. The good fit to the random cubic $H_c$ model for large grain sizes corresponds to the low nitrogen regime where the cubic anisotropy is expected to dominate. As one moves to the low $D$ part of the inset graph, the data depart drastically from the $D^6$ fit based on cubic anisotropy. This is a signature of the rising uniaxial effects. Viewed as a whole, these data show a clear transition in the coercive force mechanism in the 6 - 8 at. % nitrogen range.

Figure 6 (a) and the discussion above provided arguments for a coercive force based on cubic anisotropy for low $x_N$ values. Figure 6 (b) provides parallel arguments for a coercive force origin in the rising uniaxial anisotropy for the high $x_N$ range of compositions. This is the message given by the match up between the data and the S-W random uniaxial grain model predictions for $x_N \geq 7$ at. % or so. The validity of a cubic mechanism for lower nitrogen levels also explains the rapid rise in the S-W model prediction away from the data at low $x_N$.

This match up in Fig. 6 (b) for $x_N \geq 7$ at. %, however, raises an important question. The FMR derived and the hard direction loop based $H_u$ values shown by the solid points in (b) correspond to the overall uniaxial anisotropy for the entire film in a saturated state. The S-W model and the open points, on the other hand, are based on the *random* distribution of uniaxial grains. How can these two very different situations give a match? Most likely, the answer lies in the nature of the interactions between the random grains in the film. The same arguments used by Herzer in Refs. [31] and [32] that lead to the cubic random grain model $D^6$ coercive force response used in (a) would suggest that a large local uniaxial anisotropy is reduced substantially because of the grain to grain interactions.[34] This would lead to the type of match-up shown, as will be considered further in the next section. From an empirical point of view, this appears directly in the high dispersion in the uniaxial anisotropy evident from the open hard direction hysteresis loop in Fig. 5 (b).

## VI. ANISOTROPY ENERGY, EXCHANGE LENGTH, AND GRAIN INTERACTIONS

Section V considered FMR and hysteresis loop data and, based on these data, presented qualitative arguments on coercive force origins due to cubic anisotropy at low nitrogen levels and uniaxial anisotropy at high nitrogen levels. This section provides a more quantitative perspective on these conclusions. The analysis below is based on the classic Néel[35] and S-W[33] coercive force models for polycrystals with random cubic or uniaxial grains, respectively.

Following Harte[34] and Herzer,[31, 32] the usual cubic anisotropy energy density parameter $K_1$ is replaced by an averaged $\langle K_1 \rangle$ that takes the exchange coupling between the random cubic grains into account. The connection between $\langle K_1 \rangle$ and the local $K_1$ is related to the grain size $D$ and the so-called ferromagnetic exchange length $L_{\text{ex}}$, defined here as $\sqrt{A / \langle K_1 \rangle}$. For situations with $L_{\text{ex}} > D$, one has a $\langle K_1 \rangle / K_1$ ratio that is less than one and on the order of $(D / L_{\text{ex}})^{3/2}$. Now add coercive force considerations. As shown by Néel, the coercive force in a system of randomly oriented particles with cubic anisotropy follows the relation $H_c = 0.64 K_1 / M_s$. For interacting grains, this carries over to $H_c = 0.64 \langle K_1 \rangle / M_s$. The coercive force data presented in Sec. V, when cast into a $\langle K_1 \rangle$ vs. $x_N$ format, shows this down-scaling effect nicely. These results will be considered shortly.

Similar arguments should apply to polycrystals with a random uniaxial anisotropy. Here, one deals with the corresponding uniaxial energy density parameter $K_u = H_u M_s / 2$, but with $K_u$ replaced by $\langle K_u \rangle$ to account for the substantial anisotropy dispersion. Recall that both the open hard direction loops in Fig. 5 (b) and the $H_u$ fits to a random uniaxial anisotropy model at high nitrogen levels shown in Fig. 6 (b) support a conclusion that there are large fluctuations in the easy axis directions from grain to grain. In this case, the working formula from the classic S-W analysis for non-interacting grains with random uniaxial axes is given as $H_c = 0.96 K_u / M_s$. Interactions then give a $K_u \rightarrow \langle K_u \rangle$ replacement just as in the cubic case. The present samples, however, have a field induced rather than a random anisotropy. The interesting point is that the $H_u$ obtained from the random model, as applied to the coercive force data, matches nicely to the $H_u$ – values from the hard direction loop and FMR results. The same situation is reflected when all of the data are cast into a $K_u$ or $\langle K_u \rangle$ vs. $x_N$ format.

Figure 7 presents results on $\langle K_1 \rangle$ and $\langle K_u \rangle$ vs. $x_N$, as extracted from the coercive force and anisotropy field data in Fig. 6, based on the working relations given above. The open squares and circles show the $\langle K_1 \rangle$ and $\langle K_u \rangle$ values, respectively, as obtained from the easy direction $H_c$ data. The grain size $D$ is marked for the data points for the films with $x_N < 7$ at. %. The solid curve shows a generated line following $\langle K_1 \rangle = 15 \times 10^3 + 3.7 \times 10^{37} D^6$ and a linearized $x_N(D)$ folded in from Table I for $D \geq 10$ nm. Recall that a $D^6$ exchange coupled random grain response was verified for the $H_c$ data in Fig. 6 (a). The solid triangles and circles are from the $H_u \rightarrow \langle K_u \rangle$ conversion, based on the FMR and hard axis hysteresis loop data. The dashed straight line provides a guide to the eye for the more-or-less linear response indicated by the open circle, solid circle, and triangle data points. Note that this line extends through the origin of the graph. The shaded region shows the inferred transition region between cubic anisotropy dominant to uniaxial anisotropy dominant coercive force processes in the $6 < x_N < 8$ at. % range of nitrogen.

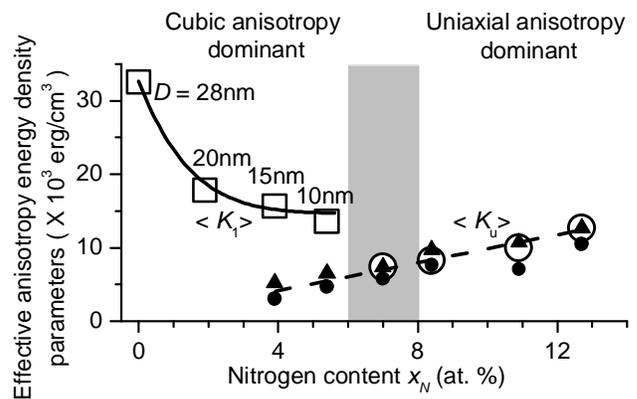

FIG. 7. Effective anisotropy energy density parameters as a function of $x_N$ at 300 K. The open squares and circles show the effective cubic and uniaxial anisotropy energy density parameters $\langle K_1 \rangle$ and $\langle K_u \rangle$, respectively, as obtained from the easy axis coercive force data. The open symbols were made bigger to give a clear visual display of all the points. The grain sizes ($D$) for low $x_N$ films are as marked. The solid curve shows a generated line of form $\langle K_1 \rangle = X' + Y' D^6$. The solid circles and triangles show the $\langle K_u \rangle$ values from the FMR and hard axis hysteresis loop data, respectively. The dashed line is drawn as a guide to the eye. The shaded region shows the possible changeover region from a dominant cubic to a dominant uniaxial anisotropy.

Figure 7 underscores three important results. First, the extracted $\langle K_1 \rangle$ values for low $x_N$ are in the range of $15-30 \times 10^3$ erg/cm$^3$ or so, much smaller than typical $K_1$ values for iron and iron alloys with low substitution levels. This order of magnitude reduction in the $\langle K_1 \rangle$ for the low $x_N$ regime is due to the exchange coupling random grain effects noted above. With $L_{ex} = \sqrt{A/\langle K_1 \rangle}$, one obtains an exchange length of about 95 nm for the $x_N = 0$ film, that is about $3-4$ times higher than the literature value for iron and iron alloys.[31, 36] However, the relation $\langle K_1 \rangle / K_1 \approx (D/L_{ex})^{3/2}$, along with the numerical data given above, gives the single grain $K_1$ value at $x_N = 0$ to be about $200 \times 10^3$ erg/cm$^3$, that is the same order as known values for iron.

Second, the fall-off in $\langle K_1 \rangle$ with $x_N$ in this same range appears to match nicely with the $D^6$ prediction from the exchange coupled random grain model. This match can also be made quantitative. The formulae given above can be combined to give a $\langle K_1 \rangle = (K_1^4 / A^3) D^6$ type response. The solid line fit in Fig. 7 corresponds to a function $\langle K_1 (\text{erg} / \text{cm}^3) \rangle = X' + Y'[D(\text{cm})]^6$, with $X' = 15 \times 10^3$ erg/cm$^3$ and $Y' = 3.7 \times 10^{37}$ erg/cm$^9$. A comparison between the $Y'$ from the fit and the $K_1^4 / A^3$ prediction gives a $K_1$ value of about $180 \times 10^3$ erg/cm$^3$, also in the right range for iron.

Finally, the coercive force data in the $x_N > 4$ at. % regime convert to $\langle K_u \rangle$ values that match the anisotropy energy density parameters extracted from the $H_u$ values from both the FMR and the hard direction loops. These values also match $\langle K_u \rangle$ estimates recently obtained from a two magnon analysis of FMR linewidth data for Fe-Ti-N films.[37, 38] One can see, moreover, that at the low end of this range, the FMR and hard direction loop data give $\langle K_u \rangle$ values that are significantly lower than the $\langle K_1 \rangle$ values inferred from the easy direction loop coercive force data. These larger $\langle K_1 \rangle$ values are the reason that the coercive force origins for the $x_N = 3.9$ at. % and $x_N = 5.4$ at. % samples lie in the cubic anisotropy, even though the uniaxial anisotropy dominates the FMR and hard axis loop responses.

The nice match up for the $\langle K_u \rangle$ values for all three types of measurements for $x_N > 7$ at. % reaffirms the dominance of uniaxial anisotropy in this region. The linear increase in $\langle K_u \rangle$ with $x_N$ also reinforces the origin of the uniaxial anisotropy in the field induced directional ordering of the interstitial site nitrogen atoms. From the dashed line in Fig. 7, $\langle K_u \rangle / x_N$ is equal to about $950 \pm 150$ erg/cm$^3$ per at. % nitrogen. van de Riet *et al.*[21] have calculated the averaged single ion uniaxial anisotropy energy for nanocrystalline Fe-Ta-N films to be $\sqrt{15 k_B T_s \langle K_u \rangle / 2N}$, where $T_s$ is the temperature of the system during the field induced ordering in-field deposition and $N$ is the concentration of the nitrogen atoms. For sputtering onto substrates at room temperature, the $\langle K_u \rangle / x_N$ ratio from Fig. 7 give an effective uniaxial anisotropy energy per nitrogen atom of about $30 \pm 10$ J/mole. This is the same value as obtained in Ref. [21] for Fe-Ta-N. Ref. [21] also cites values of 38 J/mol for Mn-Bi and 34 J/mol for Fe-C. The reasonable match for these different systems strongly supports a model for the anisotropy based on impurity atom induced structural ordering.

## VII.  SUMMARY AND CONCLUSION

In summary, the above sections have described the preparation details and measurement results on the fundamental magnetic properties of Fe-Ti-N films with nitrogen concentrations ranging from 0 to 12.7 at. % and a nominal amount of titanium at 3 at. %. The films were deposited by magnetron sputtering in an in-plane field. The focus of this work was on the effect of interstitial nitrogen on the magnetization, the exchange, the coercive force, the cubic magnetocrystalline anisotropy, and the field induced uniaxial anisotropy. The data were obtained from SQUID measurements of the magnetization vs. field and temperature and from room temperature FMR measurements.

The magnetization vs. temperature data indicate an expansion in the lattice with increasing nitrogen content and a structural transition in the $6-8$ at. % nitrogen range. A Bloch spin-wave analysis of these data to give the nearest-neighbor spin-spin exchange energy as a function of $x_N$ also indicates a lattice expansion with nitrogen below about 6 at. % and a leveling off above this level. This is also suggestive of a structural transition.

The hysteresis loop and FMR data show significant and systematic changes in the magnetic anisotropy as a function of nitrogen content. The hard direction saturation field and FMR data show the increase in the uniaxial anisotropy field with nitrogen content for $x_N > 4$ at. %, and no uniaxial character for lower nitrogen levels. The easy direction coercive force $H_c$ shows a decrease with increasing $x_N$ for nitrogen levels below about 7 at. % and then an increase for higher concentrations.

In the low nitrogen regime, $H_c$ scales with the sixth power of the grain size $D$, more or less as expected for random grains with cubic anisotropy. Extracted values of the averaged cubic anisotropy energy density parameter $\langle K_1 \rangle$ show both the expected order-of-magnitude reduction from the usual iron-based $K_1$ values and quantitative agreement with the predicted $D^6$ response.

In the high nitrogen regime, the $H_c$ values, in combination with predictions from the Stoner-Wohlfarth model, give values of the uniaxial anisotropy field $H_u$ and uniaxial anisotropy energy density parameter that are consistent with both the hard axis and the FMR data. The single ion anisotropy energies extracted from the data are also consistent with the corresponding energies for other systems, and support the origin of the anisotropy from an impurity induced structural ordering model.

## ACKNOWLEDEMENTS

Professor Chester Alexander, Jr., and Dr. Yunfei Ding of the University of Alabama, MINT Center kindly provided the Fe-Ti-N films for this research as well as extensive work on materials characterization and many useful discussions. This work was supported in part by the U. S. Office of Naval Research, Grant No. N00014-06-1-0889 , the U. S. Army Research Office, Grant No. W911NF-04-1-0247 , and the INSIC              EHDR              Program.

-----------------------------------------